# Room-Temperature Calcium Intercalation into Graphite Catalyzed by Sodium

Akira Iyo[1,a)], Shigeyuki Ishida[1], Hiroshi Eisaki[1], Hiraku Ogino[1], Kenji Kawashima[1,2]

[1] National Institute of Advanced Industrial Science and Technology (AIST), Tsukuba, Ibaraki 305-8568, Japan

[2] IMRA JAPAN Co., Ltd., Kariya, Aichi 448-8650, Japan

[a)] Corresponding author: iyo-akira@aist.go.jp

## ABSTRACT

Calcium (Ca) insertion into graphite (C) has been considered to require elevated temperatures, and its occurrence at room temperature (RT) has been regarded as highly unlikely. Here, we demonstrate that sodium (Na) catalysis enables the formation of superconducting $CaC_6$ even at RT. In mixtures of Ca, Na, and graphite, the gradual development of superconducting diamagnetism and the emergence of X-ray diffraction peaks confirm the formation of $CaC_6$ during storage at RT. The superconducting transition temperature increases with storage time, and the amount of $CaC_6$ scales proportionally to the square root of storage time. These findings provide new insights into the mechanism of superconductivity and Na-catalyzed formation of $CaC_6$, and highlight the potential of this RT intercalation process for practical applications such as electrode materials in Ca-ion batteries.

Graphite (C) has long served as a model host for studying intercalation phenomena.[1,2] While monovalent lithium ions such as $Li^+$ and $K^+$ readily intercalate into graphite to form $LiC_6$ and $KC_8$, and related compounds, the intercalation of divalent ions—such as alkaline-earth metals $AE^{2+}$ ($AE$ = Ca, Sr, Ba) and lanthanides $LN^{2+}$ ($LN$ = Sm, Eu, Yb)—is strongly hindered by electrostatic interactions with the graphite layers. Consequently, the synthesis of graphite intercalation compounds (GICs) containing divalent ions typically requires high temperatures (350–450 °C) and extended reaction times (several days).[3-9]

However, recent studies have revealed that sodium (Na) can act as a catalyst, enabling rapid intercalation of $AE^{2+}$ and $LN^{2+}$ ions at significantly lower temperatures (e.g., 250 °C for Ca) within a few hours.[10-12] The catalytic mechanism is believed to involve a transient high-stage Na-GIC ($NaC_{64}$), which lowers the kinetic barrier for intercalation and facilitates divalent-ion diffusion between graphite layers. Despite these advances, divalent-ion intercalation has still been regarded as infeasible at room temperature (RT).

Here, we investigate whether Na catalysis remains effective for the formation of $CaC_6$ at RT. $CaC_6$ is a superconductor with a transition temperature ($T_c$) of 11.5 K,[13] allowing its formation to be sensitively detected through magnetization measurements. By tracking the emergence of superconducting diamagnetism and X-ray diffraction (XRD) peaks over extended storage times, we demonstrate that Ca indeed intercalates into graphite to form $CaC_6$ even at RT. This finding offers new insights into the intercalation mechanism and opens a pathway toward practical applications of Na-catalyzed Ca intercalation into graphite.

Ca powder, graphite powder, and Na ingot were weighed in a molar ratio of 1.1:6.0:2.0 and mixed in an argon-filled glovebox using a zirconia mortar and pestle.[11] The mixture was cold-pressed into pellets (~5 mm diameter, ~1 mm thickness). A control sample without Na was prepared under the same conditions. For magnetization measurements, a pellet was sealed in a gold capsule to prevent air exposure, while pellets for XRD measurements were sealed individually in evacuated quartz tubes. All samples were stored at RT (~25 °C), and measurements were performed as a function of storage time.

Magnetization was measured using a SQUID magnetometer (MPMS, Quantum Design). Samples were first cooled to 2 K in zero field, then a magnetic field of 10 Oe was applied. Zero-field-cooled (ZFC) measurements were performed during warming to 14 K, and field-cooled (FC) measurements were subsequently performed during cooling to 2 K under the same field. XRD patterns were obtained using CuKα radiation (Rigaku Ultima IV) with airtight sample holders. The X-ray–irradiated area was standardized to allow direct comparison of diffraction intensities.

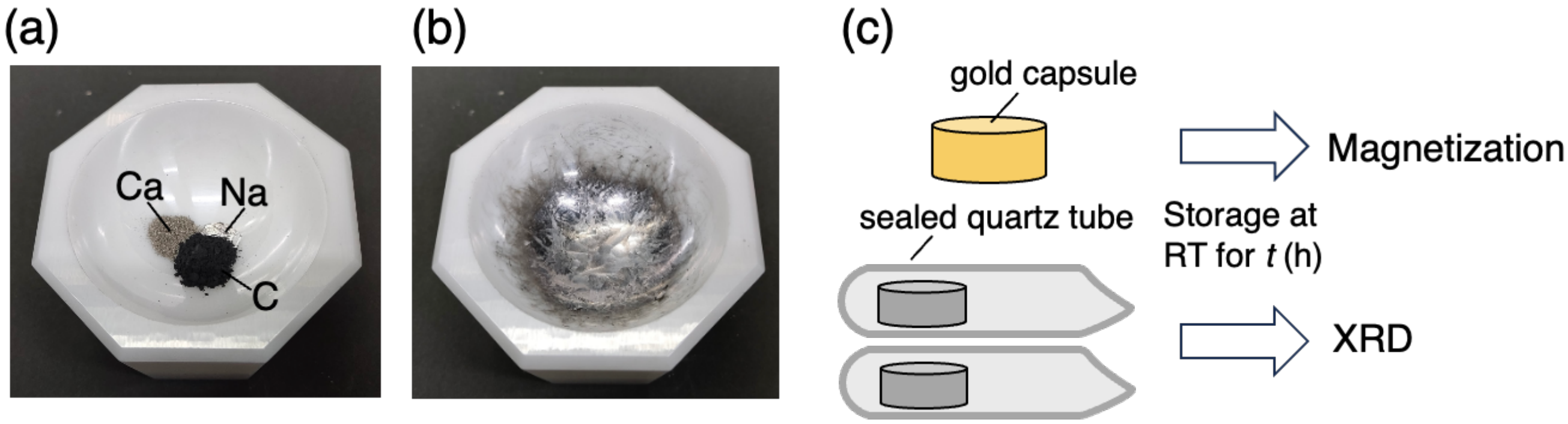

FIG. 1. Experimental workflow: (a) weighing of Ca, C, and Na, (b) mixing using a zirconia mortar and pestle, (c) sealing in gold capsule or quartz tube, and (d) magnetization and XRD measurements after storage at RT for $t$ hours.

Generally, in field-cooled (FC) measurements of superconductors, diamagnetism arising from magnetic flux expulsion (the Meissner effect) is observed. However, flux pinning significantly limits flux expulsion in type II superconductors, so the measured diamagnetism is typically small. Nevertheless, FC signals reflect the superconducting volume fraction. In contrast, zero-field-cooled (ZFC) susceptibility exhibits stronger diamagnetic signals due to magnetic shielding by superconducting regions near the sample surface.

In this study, as shown in Figs. 2(a) and 2(b), characteristic differences between the FC and ZFC susceptibilities ($4\pi M^{FC}/H$ and $4\pi M^{ZFC}/H$, respectively) are observed. Both in the FC and ZFC susceptibilities, no diamagnetic transition was detected immediately after mixing ($t$ = 0.2 h), but a clear diamagnetic signal appeared for $t$ = 6 h. The magnitude of $4\pi M^{FC}/H$ increased monotonically with $t$, clearly indicating the progressive growth of $CaC_6$ at RT. In contrast, the magnitude of $4\pi M^{ZFC}/H$ increased rapidly with $t$, reaching a nearly saturated value (≈ –0.7 at 2 K) by $t$ = 20 h. This behavior arises because the $CaC_6$ formed near the sample surface effectively shields the magnetic field, even when the overall amount of $CaC_6$ in the sample remains small.

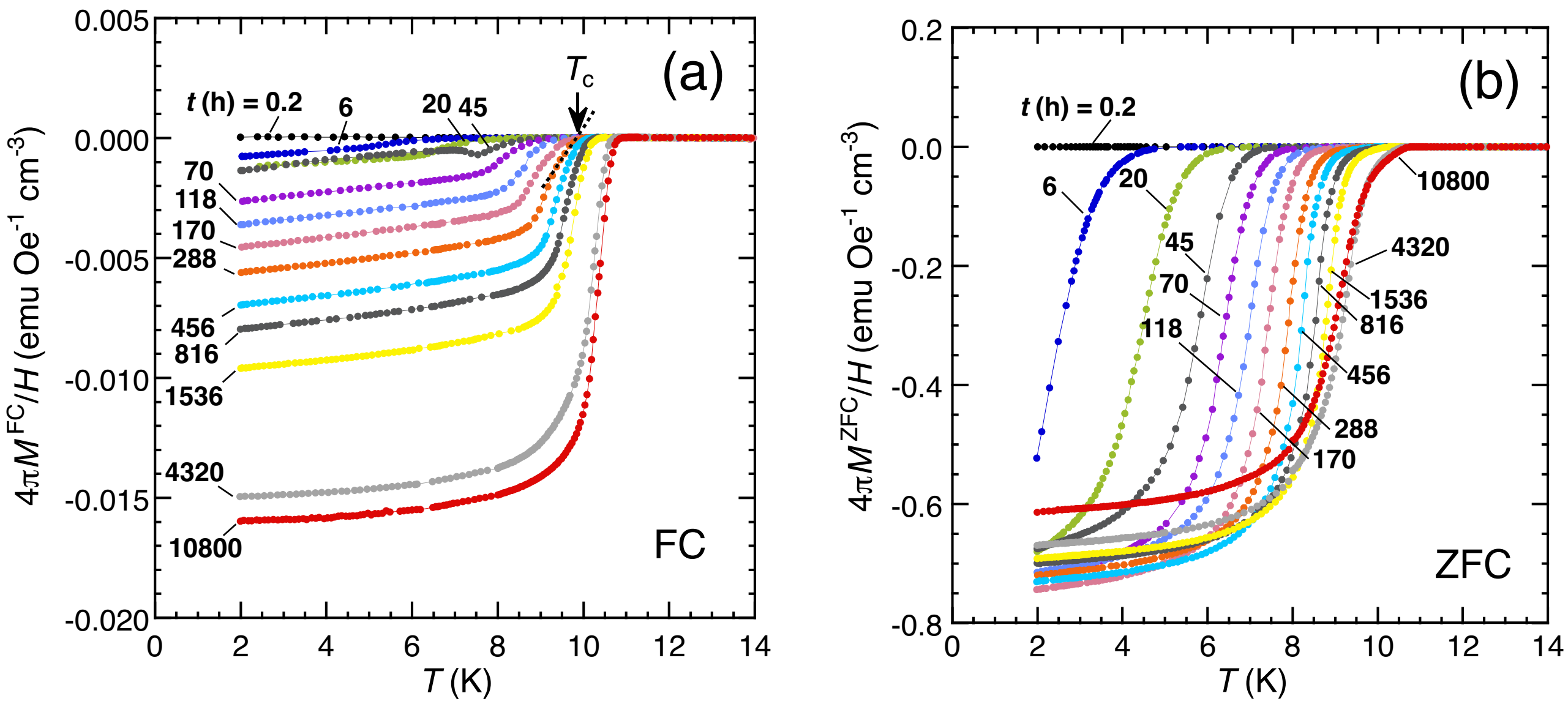

FIG. 2. Temperature dependence of (a) FC and (b) ZFC susceptibilities for samples stored for $t$ hours at RT. The data were corrected for demagnetizing effects based on the sample shape.[14]

The time-dependent formation of $CaC_6$ was directly confirmed by XRD measurements. Figure 3(a) shows XRD patterns for samples ($t$ = 48, 3120, 10800, and 24800 h), together with that of a Na-free sample ($t$ = 23800 h). For short storage times ($t$ = 48 h), diffraction peaks corresponding to $NaC_{64}$—formed by the reaction between C and Na upon mixing at RT—were observed along with peaks from unreacted Ca and Na, whereas the $CaC_6$ peak was barely detectable. With increasing $t$, the $CaC_6$ peak intensified, while the $NaC_{64}$ and Ca peaks diminished, indicating that $CaC_6$ formation progresses via a reaction between $NaC_{64}$ and Ca. In contrast, the Na-free sample exhibited no $CaC_6$ peaks even after prolonged storage ($t$ = 23800 h), demonstrating the critical catalytic role of Na at RT.

Figure 3(b) summarizes the $t$-dependence of $T_c$, defined by the onset of the FC transition (e.g., $t$ = 288 h in Fig. 2(a)). As $t$ increased, $T_c$ gradually rose, reflecting the increased electron transfer from $Ca^{2+}$ to the graphite layers as Ca intercalation progressed, eventually approaching the reported value of 11.0 K for $CaC_6$ synthesized catalytically at 250 °C.[11] This study provides the first observation of $T_c$ evolution up to its maximum value. Figure 3(c) shows the $\sqrt{t}$-dependence of $-4\pi M^{FC}/H$ at 2 K and the intensity of the strongest $CaC_6$ diffraction peak ($I_{CaC6}$ at $2\theta \approx 40°$). Both quantities, proportional to the amount of $CaC_6$ formed, scale linearly with $\sqrt{t}$, indicating that $CaC_6$ formation proceeds via a diffusion-controlled reaction.[15] These results provide new insights into the mechanism of superconductivity and the Na-catalyzed formation of $CaC_6$, warranting further detailed theoretical and experimental investigations.

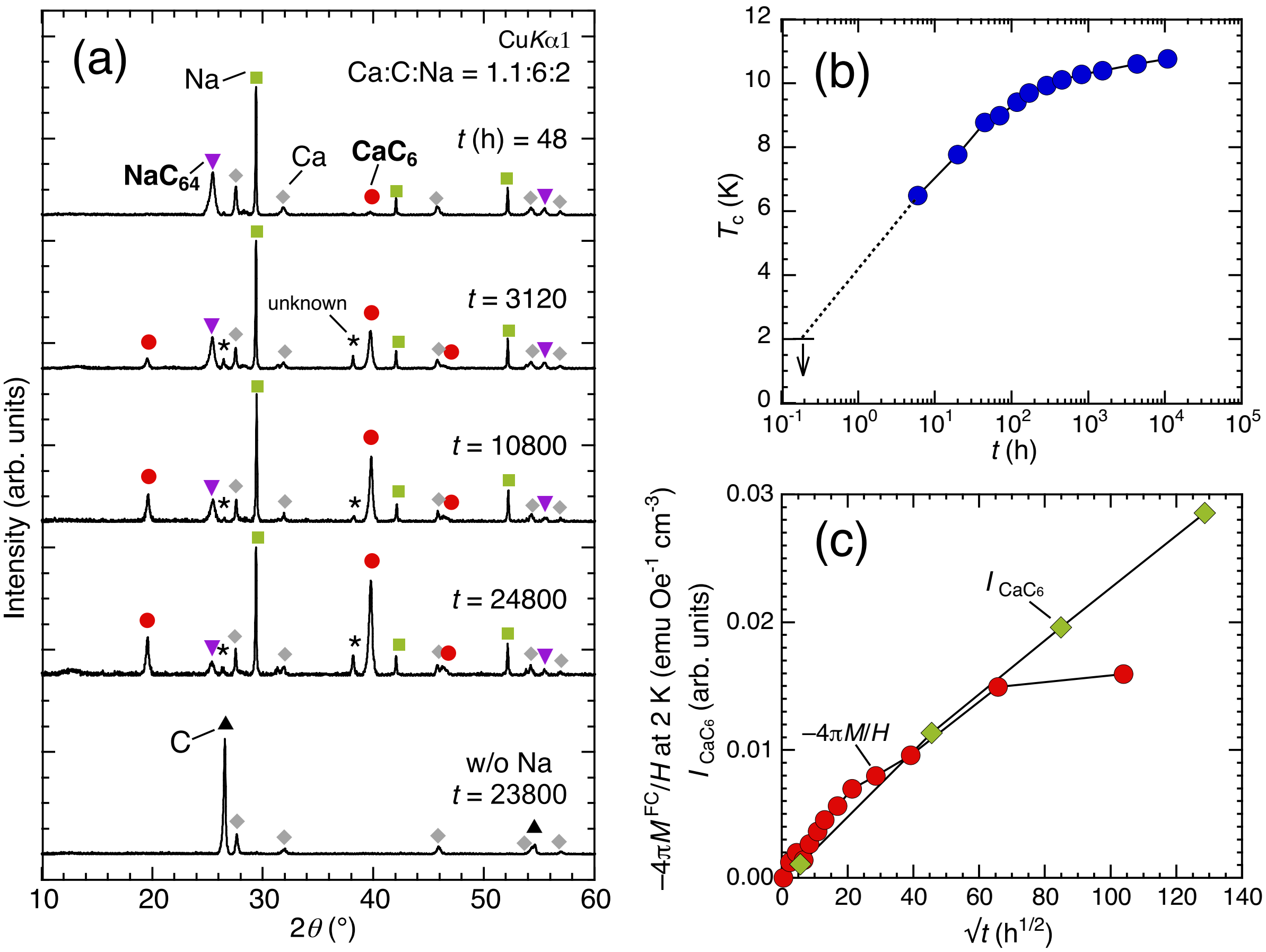

FIG 3. (a) XRD patterns after various storage times; (b) $t$-dependence of $T_c$; (c) $\sqrt{t}$-dependence of $-4\pi M^{FC}/H$ at 2 K and $CaC_6$ peak intensity $I_{CaC6}$.

This study clearly demonstrates that Ca intercalation into graphite occurs at RT through Na catalysis, expanding the potential applications of this phenomenon. One promising example is its use in rechargeable batteries. With the growing global demand for energy storage, lithium-ion batteries currently dominate the market owing to their high energy density and long cycle life.[16] However, the limited abundance and uneven geographical distribution of lithium resources, along with safety concerns, have accelerated the search for more sustainable, safer, and higher-performance alternatives.[17]

Ca-ion batteries (CIBs) are particularly attractive because Ca is both abundant and uniformly distributed in the Earth's crust, ensuring low cost and supply security. Furthermore, the divalent nature of Ca offers the potential for higher energy density compared with monovalent systems. Consequently, intensive research efforts are underway to develop practical CIBs.[18-20] A major challenge in this pursuit lies in identifying cost-effective electrode materials that enable fast and reversible (de)intercalation of $Ca^{2+}$.[21-23] Graphite stands out as a strong candidate due to its low cost and compatibility with existing LIB manufacturing infrastructure.[24,25] Our findings suggest that Na-assisted Ca intercalation can significantly enhance the electrochemical performance of graphite-based CIBs, providing valuable insights for the design of practical CIB electrode materials.

In conclusion, we have demonstrated that the catalytic effect of Na remains active for Ca intercalation even at RT, as evidenced by the gradual development of superconducting diamagnetism and the growth of diffraction peaks associated with the formation of $CaC_6$. A control sample without Na exhibited no such behavior, reaffirming the crucial catalytic role of Na. The time-dependent evolution of $T_c$ and the amount of $CaC_6$ formed were revealed for the first time, providing new insights into the mechanism of superconductivity and the Na-catalyzed formation of $CaC_6$. These findings are expected to stimulate further research on the practical utilization of the Na-catalyzed intercalation into graphite at RT, including its application for CIBs.

ACKNOWLEGEMENTS

This work was carried out in part through a collaboration between IMRA JAPAN Co., Ltd. and the National Institute of Advanced Industrial Science and Technology (AIST). This study was supported by the Japan Society for the Promotion of Science (JSPS) through Grants-in-Aid for Scientific Research (KAKENHI, Grant No. 22K04193).

AUTHOR DECLARATIONS

Conflict of Interest

The authors have no conflicts to disclose.

DATA AVAILABILITY

The data that support the findings of this study are available from the corresponding authors upon reasonable request.